# A series of magnon crystals appearing under ultrahigh magnetic fields in a kagomé antiferromagnet

R. Okuma[1], D. Nakamura[1], T. Okubo[2], A. Miyake[1], A. Matsuo[1], K. Kindo[1], M. Tokunaga[1], N. Kawashima[1], S. Takeyama[1], and Z. Hiroi[1]

**Search for a new quantum state of matter emerging in a crystal is one of recent trends in condensed matter physics. For magnetic materials, geometrical frustration and high magnetic field are two key ingredients to realize it: a conventional magnetic order is possibly destroyed by competing interactions (frustration) and is replaced by an exotic state that is characterized in terms of quasiparticles, that are magnons, and the magnetic field can control the density and chemical potential of the magnons. Here we show that a synthetic copper mineral, Cd-kapellasite, comprising a kagome lattice made of corner-sharing triangles of $Cu^{2+}$ ions carrying spin-1/2 exhibits an unprecedented series of fractional magnetization plateaux in ultrahigh magnetic fields up to 160 T, which may be interpreted as crystallizations of emergent magnons localized on the hexagon of the kagome lattice. Our observation reveals a novel type of particle physics realized in a highly frustrated magnet.**

In quantum magnets, large quantum fluctuations enhanced by frustration or low dimensionality can realize non-trivial phases of matter. One of the examples is the quantum spin liquid with spins fluctuating coherently down to absolute zero temperature[1,2]. It has macroscopic entanglements which support exotic fractionalized excitations such as anyons or majorana fermions[1,2]. The spin-1/2 Heisenberg kagome antiferromagnet (KAFM) here we focus is believed to be a candidate for the quantum spin liquid.[15-17] Another example is found in the spin-1 Heisenberg chain, which is characterized by a non-trivial gap and a spin-1/2 moment emergent at a domain edge[3]. These quantum phases of matter are now categorized as the symmetry-protected topological phase, which has an emergent state in a lower dimension, such as a surface state in a topological insulator[4].

Magnetic field is another powerful tool to realize nontrivial quantum phases because it can directly affect a spin state through Zeeman coupling. In a paramagnetic state it forces spins to precess around its direction with a Larmor frequency. The phase of this circular motion is similar to the phase factor of the wave function of bosons or fermions. In fact, numerous cooperative phenomena in quantum spin systems with the global rotational symmetry can be described in terms of strongly correlated hard-core bosons called magnons, which are similar to the typical bosonic particles of $^4He$[5].

Let the vacuum of magnons to be a state without symmetry breaking: here we consider a singlet state in a dimer magnet or a fully polarized state in a Heisenberg antiferromagnet that may exhibit a long-range order (LRO) at zero field. Elementary magnon excitations from these states are the formations of one triplet in the sea of singlets and one down spin in the sea of up spins, respectively. In both cases, magnetic field acts as a chemical potential for magnons and thus controls the density of magnons, which is proportional to the magnetization. Magnons are mobile and tend to exhibit a Bose Einstein condensation (BEC) as a liquid condenses into a superfluid state. Field-induced transverse magnetic order in the dimer system[6] and the appearance of LRO just below the magnetization saturation[7] are interpreted as BECs of magnons.

Magnons tend to localize in some frustrated magnets owing to the suppression of kinetic energy and eventually crystallize to become "insulating", as in Mott insulators in strongly correlated electron systems. This magnon crystal has a fixed density of magnons, so that the magnetization remains at a fractional value of the full magnetization in a range of field in the magnetization process, which is called the magnetization plateau[8]. The fractional value of magnetization comes from the commensurability of the magnon crystal: the number of magnons, $Q_{mag} S(1 - m)$, in the magnetic unit cell should be an integer, where $Q_{mag}$ is the number of spins in the magnetic unit cell, $S$ is the spin quantum number, and $m$ is the magnetization divided by the saturation magnetization[9,10].

Magnetization plateaux have been discovered in various frustrated quantum magnets[8]. $SrCu_2(BO_3)_2$ is a typical example, which comprises pairs of $Cu^{2+}$ ions arranged orthogonally to each other in the sheet and possesses a Shastry–Sutherland state made of spin-1/2 dimers on the pairs with a spin gap[11]. At magnetic fields larger than the gap, a series of magnetization plateaux appear at $m = 1/8, 1/4, 1/3$ ($Q_{mag} = 16, 8, 12$)[12,13]; NMR measurements directly confirmed spontaneous translational symmetry breaking in these magnon crystals[14].

[1]Institute for Solid State Physics, University of Tokyo, Kashiwa, Chiba 277-8581, Japan. [2]Department of Physics, The University of Tokyo, Tokyo 113-0033, Japan.  e-mail: rokuma@issp.u-tokyo.ac.jp

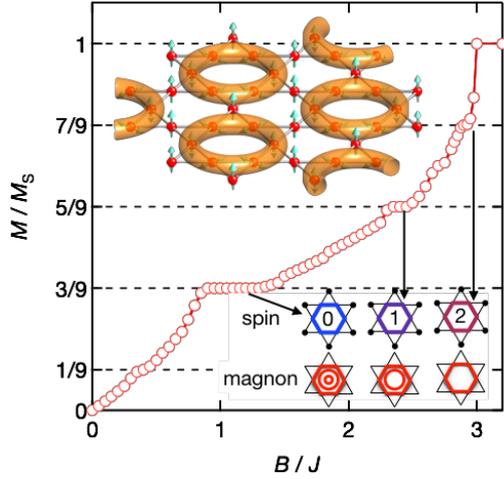

**Fig. 1 | Calculated magnetization process for the spin-1/2 KAFM with the nearest-neighbor interaction $J$.** The tensor network method with the Projected Entangled Pair State (PEPS) is used. The vertical and horizontal axes represent magnetization $M$ divided by saturated magnetization $M_s$ and magnetic field $H$ divided by $J$, respectively. The top left inset shows a schematic drawing of hexagonal magnons that are depicted by doughnuts containing six entangled spins. The other intervening spins point up in the direction of magnetic field. The magnon crystal forms a superlattice with a $\sqrt{3} \times \sqrt{3}$ unit cell. The bottom right inset shows hexagonal magnons expected to appear at the 1/3, 5/9 and 7/9 plateaux. In the upper part, the magnons are defined by the total spin $S^z = 0$, 1 and 2 for the 6 spins on the hexagon, respectively, while, in the lower part based on the magnon picture, the number of magnons are 3 (hexagon + double circle), 2 (single circle) and 1 (only hexagon), respectively.

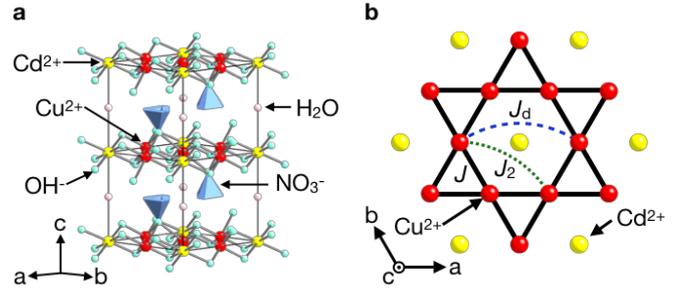

**Figure 2 | Crystal structure and the magnetic interactions of CdK. a**, kagomé layers comprising $Cu^{2+}$ and $Cd^{2+}$ ions possess good two dimensionality, separated by nonmagnetic nitrate ions and crystalline water molecules. **b**. regular kagome lattice made of spin 1/2 $Cu^{2+}$ ions with a Cd ion in the center of each hexagon. Magnetic interactions between spins are nearest-neighbor $J$, a DM interaction as large as 10% of $J$, small next-nearest-neighbor interaction $J_2$ and diagonal interaction $J_d$ across the Cd ion in the hexagon.

For the $S = 1/2$ KAFM[15-17], the formation of a nontrivial magnon is theoretically expected just below the saturation[18]. Starting from the fully spin polarized state above the saturation field $B_s$, which is the vacuum of magnons, one magnon with total $S^z = 2$ in a hexagonal plaquette among other hexagons with $S^z = 3$, as schematically depicted in Fig. 1 is generated when the magnetic field is set infinitesimally smaller than $B_s$. As each spin inside a hexagonal plaquette equally carries fractional magnetization, this "hexagonal magnon" is a highly quantum mechanical entity. Because of no energy cost for this magnon generation, the density is quickly raised to 1/9 before magnons overlap with each other to feel mutual repulsion, which results in a jump of magnetization from 1 to 7/9 at $B_s$[18]. Then, a crystalline phase with a superstructure of the $\sqrt{3} \times \sqrt{3}$ unit cell with $Q_{mag} = 9$ is formed in a range of field, giving a 7/9 magnetization plateau. Substantially different from the dimer singlet state naturally formed on a pair of Cu ions in the compounds mentioned above, a larger magnon is emergently generated on a hexagon of the kagome lattice in the KAFM.

Recent calculations by the density-matrix-renormalization-group method, the exact diagonalization, and the tensor network method show that additional plateaux appear at $m = 5/9, 1/3, 1/9$.[19-22] It is also suggested[19-21] that the 5/9 and 1/3 plateaus are magnon crystals with $Q_{mag} = 9$, which are similar to that at $m = 7/9$ but with $S^z = 1$ and 0, respectively, as depicted in Fig. 1; the 1/9 plateau is supposed to be another state, possibly a spin liquid with topological order[19]. However, one has to be careful to conclude these because there are many competing phases. In the present study, we have calculated the whole magnetization process in the extended PEPS scheme, which essentially agrees with the previous results, as shown in Fig. 1. In addition, we have carefully examined the magnetic structure at the 1/3 plateau and found that a magnon crystal made of hexagonal magnons with $S^z = 0$ in the $\sqrt{3} \times \sqrt{3}$ structure takes lowest energy among other competing states such as the up-up-down state (Supplement S1). Thus, the hexagonal magnon must be an entity to be realized under magnetic field for the KAFM, which is to be experimentally evidenced. Emergence of a new composite degree of freedom is common to highly frustrated magnets; eg. a molecular excitation in the pyrochlore antiferromagnet $ZnCr_2O_4$[23] and 10-spin loop order in the hyperkagomé antiferromaget $Gd_3Ga_5O_{12}$[24]. These new types of excitations do not only offer an interesting subject in condensed matter physics, but also would lead to novel future application when they are manipulated appropriately.

Despite the intriguing predictions for the KAFM in magnetic fields, little experimental evidence has been accumulated because of the lack of ideal model compounds and also difficulty in experiments under high fields. Real materials always suffer from lattice distortion[25] or disorder[26], which tends to mask the intrinsic magnetism of the KAFM. The most well characterized $S = 1/2$ KAFM is herbertsmithite with a large interaction of $J \sim 200$ K[27], which means that ultrahigh magnetic fields no less than $B_s = 400$ T are necessary to measure the whole magnetization process. However, experimentally available static magnetic fields are only below 45 T and typical pulsed magnetic fields are limited to below 100T[28]. Moreover, magnetization measurements by the conventional induction method are available in pulsed magnetic fields below 100 T in the short time duration of several msecs[25]. For larger pulsed fields, a much shorter time window of several μsecs causes a large electromagnetic noise that prevents facile magnetization measurements. A state-of-the-art measurement using a Faraday rotation technique has made it possible to record the whole magnetization process of $CdCr_2O_4$ up to 140 T, which reveals a spin nematic phase just below the saturation[29]. In order to unveil the physics of the KAFM, both a suitable model compound with a relatively low saturation field around 100 T and an improvement of the Faraday rotation technique are necessary; it is noted that for a compound with smaller $J$ and $B_s$ values the required low temperature condition, $T \ll J/k_B$, becomes difficult to achieve.

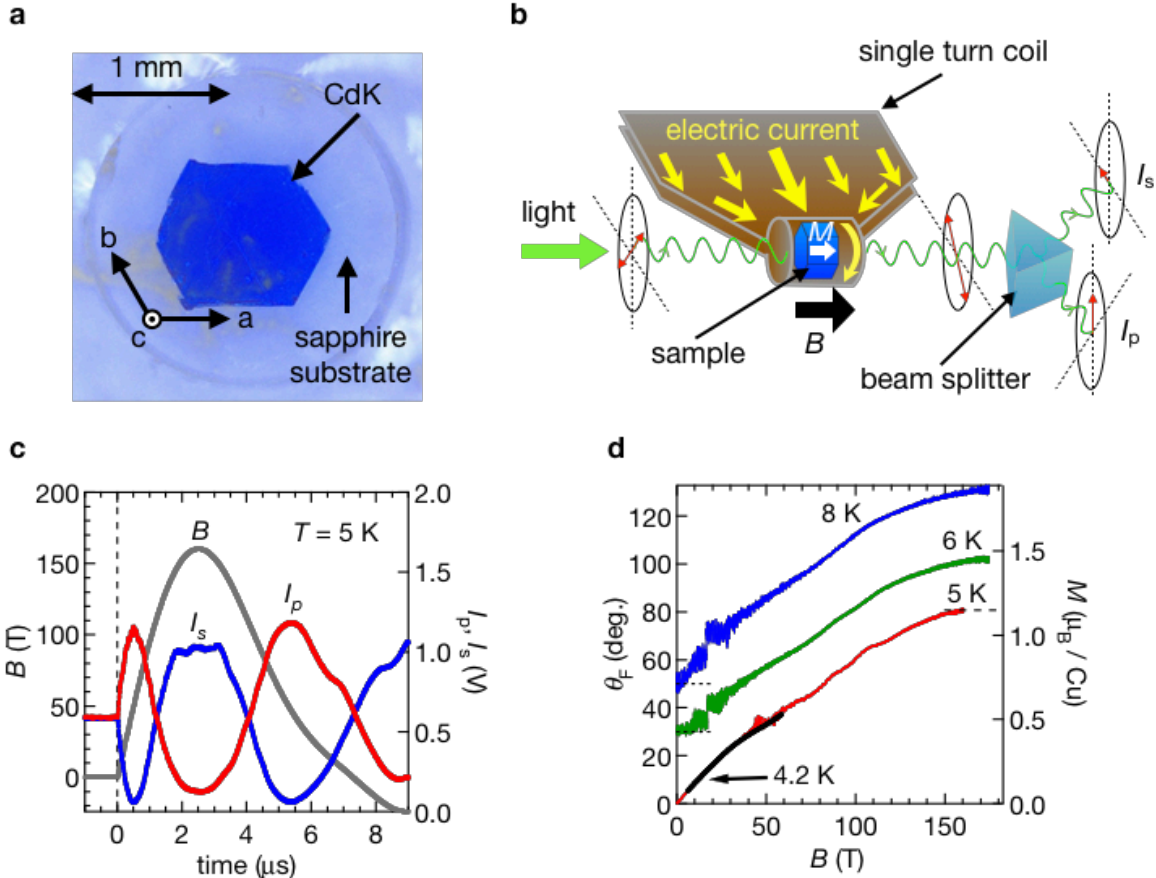

**Figure 3 | Magnetization measurements by the Faraday rotation technique in pulsed magnetic fields up to 160 T. a**, single crystal of CdK on a sapphire disk substrate used in high-field Faraday rotation measurements. The crystal has a flat (001) surface, which effectively reduces scattering of light, and is large enough to obtain a Faraday rotation signal. **b**, schematic image of the experimental setup. A high magnetic field is generated inside a single turn coil of 12 mm in inner-diameter by injecting a huge electric current (2-3 mega-ampère). Shown in **c** is a typical profile of a pulsed magnetic field with the maximum at 160 T and the time duration of 7 μsecs. The hexagonal plate-like single crystal of CdK shown in **a** mounted on a sapphire substrate is placed in the coil in such a way that magnetic field is perpendicular to the hexagonal surface ($B \parallel c$). Linearly polarized incident light comes from left and passes through the crystal. A change in the polarization angle due to the Faraday effect in the magnetized crystal is measured: transmitted light is split into vertical and horizontal components, $I_p$ and $I_s$, and each intensity is recorded by an oscilloscope. **c**, time evolutions of magnetic field, $I_p$, and $I_s$ at 5 K. **d**, field dependences of the Faraday rotation angle $\theta_F$ and the corresponding magnetizations at 5 K (red), 6 K (green) and 8 K (blue). For clarity, offsets of 20 and 50 degrees are added to the data at 6 and 8 K, respectively. The magnetization data from the Faraday rotation angle have been calibrated so as to reproduce those obtained by the induction method using a nondestructive pulse magnet as shown by the black line below 55 T at 4.2 K. The large noises at ~30 T in the data at 6 and 8 K and at ~50 T in the data at 5 K are due to reductions of intensity for either $I_p$ or $I_s$ at around the rotation angle of 90 degrees.

To study the magnetism of the $S = 1/2$ KAFM under magnetic fields, we focus on Cd-kapellasite (CdK). This compound has a quasi-two-dimensional structure with an undistorted kagome lattice of $Cu^{2+}$ ions and a moderate antiferromagnetic interaction of $J \sim 45$ K ($B_s \sim 100$ T) (Fig. 2a)[29, 30]. The ground state of CdK is not a spin liquid but a LRO with a **q** = 0 structure (a negative vector chirality order) with $T_N \sim 4$ K, which may be induced by a Dzyaloshinskii–Moriya (DM) interaction of the magnitude of approximately 10% of $J$. Other possible perturbations are the next-nearest-neighbor interaction $J_2$ and the diagonal interaction $J_d$ bridged via the Cd ion in the center of the hexagon, as shown in Fig. 2b, both of which turn out to be small (Supplement S2). A high-quality single crystal with a large, flat surface suitable for optical measurements has been obtained (Fig. 3a). We have measured the small magnetization from one tiny hexagonal single crystal of CdK (~1 mm in edge and 150 μm in thickness) by a Faraday rotation technique optimized to the single-turn coil megagauss generator in magnetic fields of up to 160 T. The experimental setup and results are shown in Fig. 3b. The polarization angle of an incident light with λ = 532 nm (Supplement S3) rotates with varying magnetic field in several μsecs by the Faraday effect from induced magnetization in the sample (Fig. 3c). Figure 3d shows magnetizations converted from the data in descending pulsed magnetic fields (Supplementary S4). The magnetization curve measured at 8 K smoothly increases and saturates at around 150 T. At 6 K, a faint wiggling is noticed and turns out to be some anomalies at 5 K. Fig. 4a shows the magnetization process at 5 K and its field derivative compared with the calculated magnetization process in the presence of DM interaction. Below 0.4 $\mu_B$, the experimental magnetization agrees quite well with the calculation. Interestingly, at least seven anomalies are observed at magnetizations greater than 0.4 $\mu_B$ while there is no magnetization plateau or anomaly in the calculated curve. Above $B_s = 160$ T, $M$ reaches 1.15 $\mu_B$/Cu, which is nearly equal to the saturation magnetization with fully polarized spins: $gS\mu_B$ with $g_c \sim 2.3$ from magnetic susceptibility (Supplementary S2).

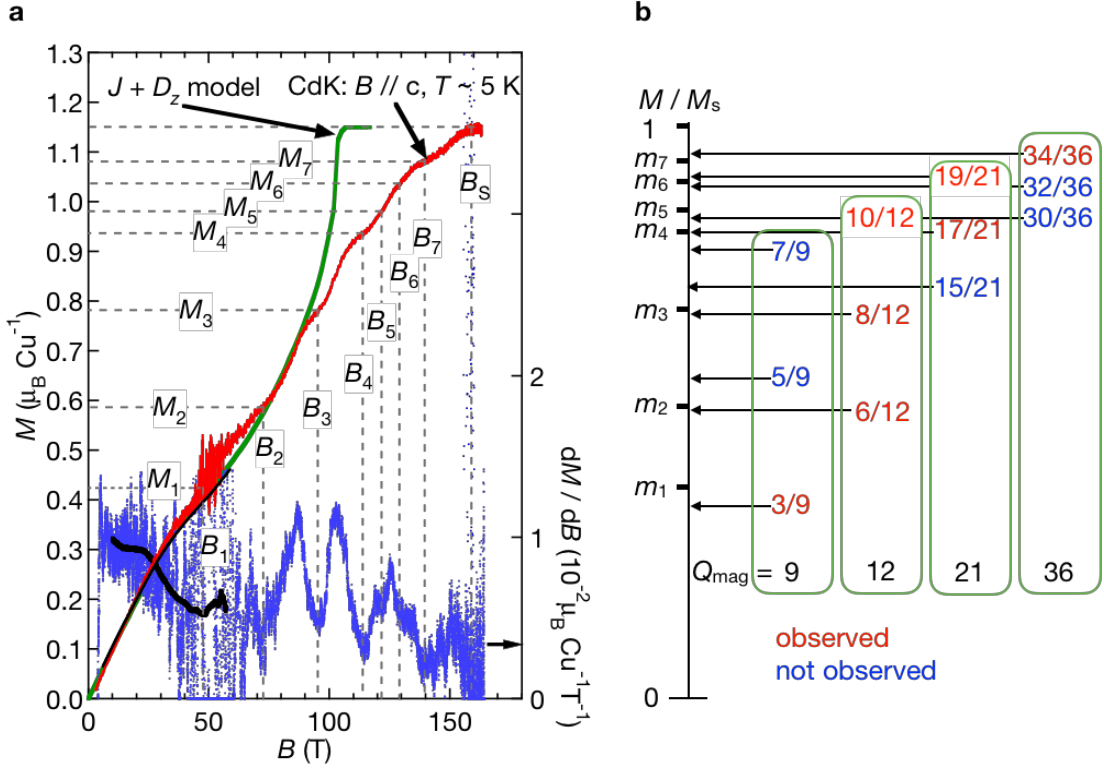

**Figure 4 | Appearance of multiple magnetization plateaux in CdK. a**, field dependences of magnetization and its derivative at 5 K and $B // c$. The red and black lines represent magnetizations measured by the Faraday rotation and the induction method, respectively, and blue points represent the field derivative of the former curve. The magnitudes of $B$ and $M$ at 7 local minima in the field derivative curve are named as $B_k$ and $M_k$ as shown by the broken lines, which may correspond to a series of magnetization plateaux. The saturation field $B_s = 160$ T is determined at which $M$ reaches the full magnetization $M_s = 1.15\mu_B$. The green line shows a calculated $M$ by PEPS for an $S = 1/2$ KAFM model with $J = 45$ K and the $z$ component of DM interaction of $0.1J$. **b**, comparison between experimentally observed plateau magnetizations $m_k = M_k/M_s$ and fractional magnetizations expected from the series of magnon crystals with the $Q_{mag} = 9, 12, 21, 36$ unit cells shown in Fig. 5. The fractions in the red letter are close to one of the $m_k$ values, while blue ones may not be observed except for 30/36 which is assigned to 10/12.

We consider these anomalies as a series of blunt plateaux that are blurred or inclined owing to the finite temperature effect or anisotropy such as DM interaction; although conventional ordered antiferromagnets can experience a metamagnetic transition or a spin-flop transition in magnetic fields, weak anisotropy in $Cu^{2+}$ spin systems must make it difficult for such transitions to take place. Taking account of blunting by temperature, we define a critical field $B_k$ and a magnetization $M_k$ for each plateau at which the differential magnetization takes a local minimum; the $B_k$ should correspond to the center of the field range for the plateau. The obtained values of [$B_k$ (T), $M_k$ ($\mu_B$/Cu), $m_k$] ($m_k = M_k/1.15$) are (47.6, 0.42, 0.37), (72.6, 0.59, 0.51), (95.3, 0.78, 0.68), (113.8, 0.94, 0.81), (121.6, 0.98, 0.85), (129.1, 1.04, 0.9), (139.7, 1.08, 0.94) for $k = 1–7$, respectively. Note that the anomalies are distinct for $k = 2, 3, 4$ and 7, while are weaker for $k = 5$ and 6. As mentioned above, it is expected for the simple K3AFM that magnetization plateaux from magnon crystals with $Q_{mag} = 9$ appear only at $m = 1/3, 5/9$ and 7/9. These values are near to $m_1, m_2$ and $m_4$, as compared in Fig. 4b, but are significantly deviated from them. The present observation of more plateaux in CdK strongly indicates the formation of other types of magnon crystals with unit cells larger than $Q_{mag} = 9$.

Let us consider possible magnetic structures realized at the magnetization plateaux in CdK. As in the case of the simple KAFM, we assume that one kind of localized hexagonal magnons are periodically aligned to reduce mutual repulsion with keeping a six-fold rotational symmetry. Then, such series of magnon crystals as shown in Fig. 5 with the unit cells of $Q_{mag} = 9, 12, 21, 36$ are possible to appear each with 3 sequences having 1–3 magnons on the hexagon. All the experimentally observed $m_k$ values are well reproduced with them. For example, $m_7 = 0.94$ near the saturation is close $34/36 \sim 0.944$, and $m_6 = 0.90$ is nearly equal to $19/21 \sim 0.905$. Moreover, $m_5 = 0.85, m_4 = 0.82, m_3 = 0.68, m_2 = 0.51$ and $m_1 = 0.37$ are near to $5/6 \sim 0.833, 17/21 \sim 0.810, 2/3 \sim 0.667, 1/2 = 0.5$ and $1/3 \sim 0.333$, respectively (Fig. 4b). As a result, one from the $Q_{mag} = 9$ series, three from $Q_{mag} = 12$, two from $Q_{mag} = 21$ and one from $Q_{mag} = 36$ are observed. The fact that all the 3 sequences appear for $Q_{mag} = 12$ whereas only part for the others prompts us to find a rule for the magnon crystallization in CdK.

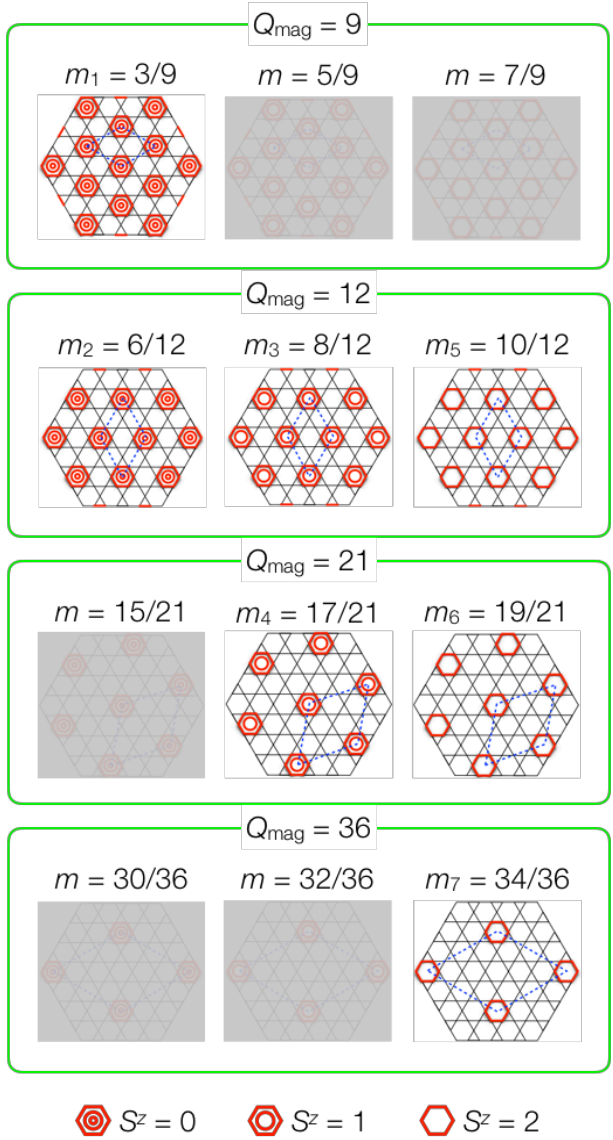

**Figure 5 | Possible series of magnetic structures for the magnon crystals with $Q_{mag}$ = 9, 12, 21 and 36.** Among them, those corresponding to the observed 7 magnetization plateaus are shown emphasized. Red hexagons represent localized hexagonal magnons with the total $S_z$ = 0, 1 and 2. Other intervening spins are fully polarized along magnetic field as shown for $Q_{mag}$ = 9 in the inset of Fig. 1. Blue dotted lozenges represent the magnetic unit cells.

Emergence of larger unit cells in CdK must be attributed to additional further-neighbor interactions such as $J_2$ and $J_d$. In fact, our calculations of magnetization without them and with 10% DM interaction show no anomalies after an inclined 1/3 plateau with increasing field (Fig. 4a). Note that the anomalies in CdK appear after the experimental magnetization curve starts to deviate from the theoretical one, which suggests that effects excluded in the calculation play a crucial role. If $J_2$ works as an effective repulsion among magnons, the $Q_{mag}$ = 9 structure becomes unstable (Supplementary S5). For the $Q_{mag}$ = 12 structure, $J_2$ does not work but $J_d$ does. The experimental fact that all the 3 sequences from the $Q_{mag}$ = 12 series are observed and only part for the others (e.g. not 5/9 but 6/12 for $m_2$) indicates the relative stability of $Q_{mag}$ = 12. This means repulsive (ferromagnetic) $J_2$ and attractive (antiferromagnetic) $J_d$. The larger unit cells like $Q_{mag}$ = 21 and 36 may be stabilized by weak further-neighbor interactions. Therefore, the appearance of multiple magnetization plateaus in CdK is reasonably explained by assuming a series of magnon crystals in a spin 1/2 KAFM with further-neighbor interactions.

Finally, we turn our attention to the reason why the magnetization plateaus in CdK appear after the deviation from the calculation based on the $J$ + DM model. The smooth rise of the calculated curve indicates that spins are gradually forced to align along magnetic field from the coplanar **q** = 0 structure, as expected for a classical spin system. However, we should remember that lattice distortion could lift classical degeneracy of frustrated systems besides anisotropy[31]. For example, a hexagon where a localized magnon lives can shrink so as to increase antiferromagnetic $J$ and thus to stabilize a local magnon. When such a lattice distortion occurs periodically to form a lattice, hexagonal magnons can crystallize at high fields with $m \geq 1/3$. Possibly for CdK, a spin-lattice coupling makes a tilted **q** = 0 magnetic structure unfavorable and leads to crystallizations of localized magnons with the help of further-neighbor interactions.

Our observation of a series of fractional magnetization plateaus in the kagome antiferromagnet CdK demonstrate the presence of emergent hexagonal magnons in the kagome lattice, which has been inferred by theoretical calculations but has never been evidenced experimentally. Moreover, they crystallize with large unit cells owing to lattice commensurability, additional long-range interactions and possibly coupling to lattice. In order to understand field-induced quantum states in KAFM in more detail, it is necessary to perform more experiments for other kagome compounds. We are now challenging a magnetization measurement at extremely high fields of 600 T, which would make it possible to record the whole magnetization process of the more ideal herbertsmithite with a large saturation field of ~400 T. We think that these results will offer not only new physics in frustrated magnetism but also new insights into elementary excitation and many-body interactions in solids.

## Methods
**Sample preparation.** Single crystals of Cd-kapellasite, $CdCu_3(OH)_6(NO_3)_2 \cdot H_2O$, were synthesized by a two-step hydrothermal method[30]. A quartz ampoule 150 mm in length was filled with ingredients ($Cu(OH)_2$: 0.1 g, $Cd(NO_3)_2 \cdot 4H_2O$: 5 g, $H_2O$: 4 g) and sealed; a thick quartz tube 12 mm in outer and 8 mm in inner diameter was used to avoid bursting due to increased pressure inside the tube during heat treatment. The ampoule was placed horizontally in a two-zone furnace and heated to 180 °C at the hot end and to 130 °C at the cold end for a week. Polycrystalline samples of CdK were quickly produced over the tube and then slowly transported into the cold zone to condense into a bunch of single crystals 100 μm in diameter. After all polycrystalline samples were transported to the cold zone, the temperature gradient was reversed: the cool zone was set to 160 °C and the hot zone to 140 °C. This caused the inverse transportation, resulting in the growth of hexagonal plate-like crystals as large as 1 mm in diameter and 150 μm in thickness, typically as shown in Fig. 3a.

**Magnetization measurements.** Magnetization measurements in low magnetic fields below 7 T were conducted using a single crystal of CdK of 2.38 mg weight in a Magnetic-Property-Measurement-System 3 (MPMS-3, Quantum Design). Magnetic susceptibilities along the **a** and **c** axes were measured in the constant field of 1 T. Magnetization at pulsed magnetic fields up to 60 T was measured at 4.2 K on stacked single crystals by the electro-magnetic induction method. The absolute value of magnetization was calibrated by the data obtained from MPMS-3.

**Faraday rotation measurements in pulsed magnetic fields.** A single-turn coil as shown in Fig. 3a was used to generate ultrahigh magnetic fields up to 180 T[32]. A sample was cooled down to 5 K and was not destroyed after experiment because the coil explodes outward along the direction of Maxwell stress. Faraday rotation was measured in a short time duration of 7 μsec. A change in the polarization angle $\theta_F$ is proportional to magnetization $M$ induced by the applied field in the sample: $\theta_F = \alpha M d$ where $\alpha$ is the Verdet constant and $d$ is the thickness of the sample. $\theta_F$ is calculated from the intensities of the vertical and horizontal components, $I_p$ and $I_s$, by the formula: $\theta_F = \cos^{-1}\{(I_p - I_s)/(I_p + I_s)\}$.

**Tensor network calculations.** Magnetization process of a $S = 1/2$ KAFM with the nearest-neighbor interaction $J$ is calculated by the infinite Projected Entangled Pair State (iPEPS)[33-35] method which expresses a wave function as an extended tensor product state so called projected simplex pair state (PESS)[36,37]. Unit cells up to 18 sites are assumed. The wave function is optimized by the simple update method[38] with the bond dimensions $D = 4$-$7$. Physical quantities were calculated by the corner transfer method[39,40], with the bond dimensions typically $\chi = D^2$.

## Acknowledgments

We are grateful to T. Misawa, H. Tsunetsugu, and C. Hotta for helpful discussion. R. O. is supported by the Materials Education Program for the Future Leaders in Research, Industry, and Technology (MERIT) given by the Ministry of Education, Culture, Sports, Science and Technology of Japan (MEXT). This work was partially supported by KAKENHI (Grant No. 15K17701), the Core-to-Core Program for Advanced Research Networks given by the Japan Society for the Promotion of Science (JSPS), and MEXT of Japan as a social and scientific priority issue (Creation of new functional devices and high-performance materials to support next- generation industries; CDMSI) to be tackled by using post-K computer.

This is a pre-print of an article published in Nature Communications. The final authenticated version is available online at: https://doi.org/10.1038/s41467-019-09063-7

## Author contribution

R. O., D. N., S. T., and Z. H. conceived and designed the study. A. Miyake, and A. Matsuo measured the magnetization curve up to 60 T under supervision of M. T. and K. K . R. O., D. N., measured the magnetization curve up to 180 T under supervision of S. T. R. O., D. N., S. T., and Z. H. interpreted the experimental data. T. O. performed and interpreted PEPS calculations under supervision of N. K. R. O. wrote the paper. All authors discussed and commented on the manuscript.

## Additional information

Supplementary information is available in the online version of the paper. Reprints and permissions information is available online at www.nature.com/reprints. Correspondence and requests for materials should be addressed to R. O.

## Competing financial interests

The authors declare no competing financial interests.

# Supplementary

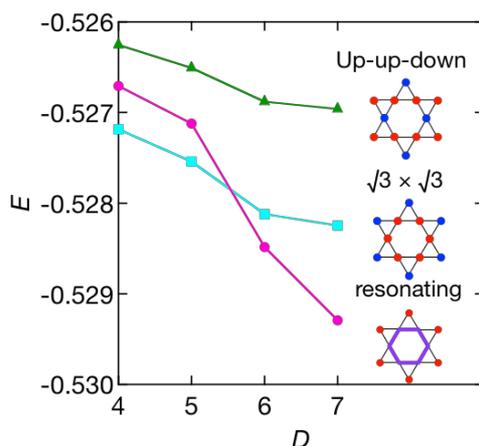

**S1 Energies at $h = 1.1J$ calculated by the PEPS method as a function of the bond dimension D for the three spin configurations expected at the 1/3 plateau.** The red and blue circles represent up and down spins, respectively, and the purple hexagon represents entangled 6 spins in the singlet state. In the up-up-down and $\sqrt{3} \times \sqrt{3}$ states, each triangle has up, up and down spins; the triangles are arranged with **q** = 0 and $\sqrt{3} \times \sqrt{3}$, respectively, while the resonating state contains a singlet hexagonal magnon. Among them, the resonating state finally yields lowest energy as the wavefunction is optimized with increasing the bond dimension $D$.

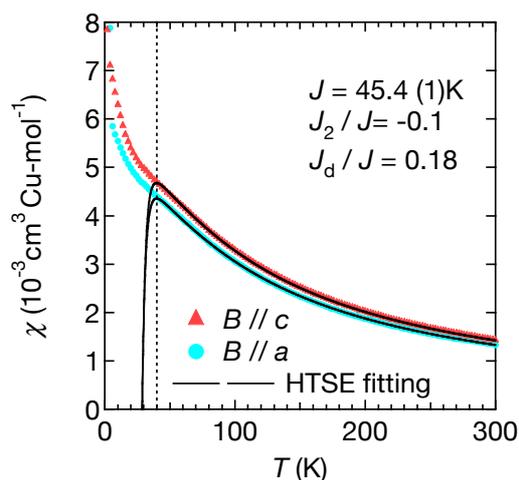

**S2 Magnetic susceptibility of CdK measured at $B = 1$ T along the $a$ (blue circle) and $c$ (red triangle) axes.** To estimate the magnitudes of $J$, $J_2$ and $J_d$, both data above 40 K are simultaneously fitted to the 10th-order-high-temperature-series expansion[41]. The best fit is obtained when $(J/K, J_2/J, J_d/J) = (45.4, -0.1, 0.18)$ with $g_a = 2.28$ and $g_c = 2.37$ and a common temperature-independent term $c_0 = -6.65 \times 10^{-5}$ cm$^{-3}$ Cu-mol$^{-1}$. Note that these values are rough estimates because they heavily depend on the temperature region used; the available information from the data is too small to deduce reliable parameters.

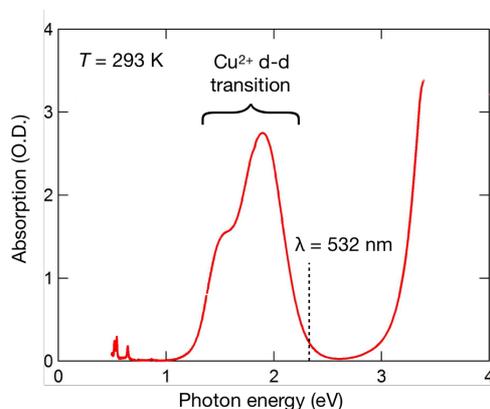

**S3 Optical absorption spectrum of CdK at room temperature.** The peak at 1.5–2 eV corresponds to the $d$-$d$ transition ($T_2 \to E$) of Cu$^{2+}$ [42]. For Faraday rotation measurements, light with $\lambda = 532$ nm from a Nd:YAG laser is employed, which locates at the bottom of the $d$-$d$ transition so that non-linear effects on the Faraday rotation can be ignored.

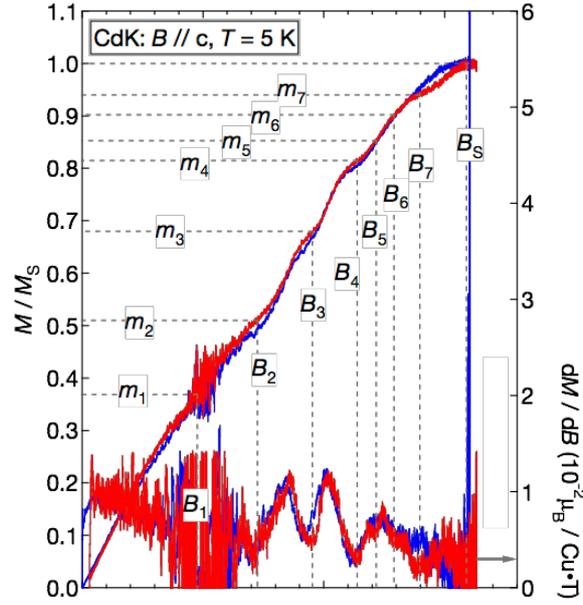

**S4 Comparison of two successive magnetization measurements with elevating (blue curves) and descending (red curves) magnetic field at 5 K.** Anomalies due to plateaus are observed at almost the same magnetic fields in the two curves. We analyze field-descending data in detail because more anomalies are observed.

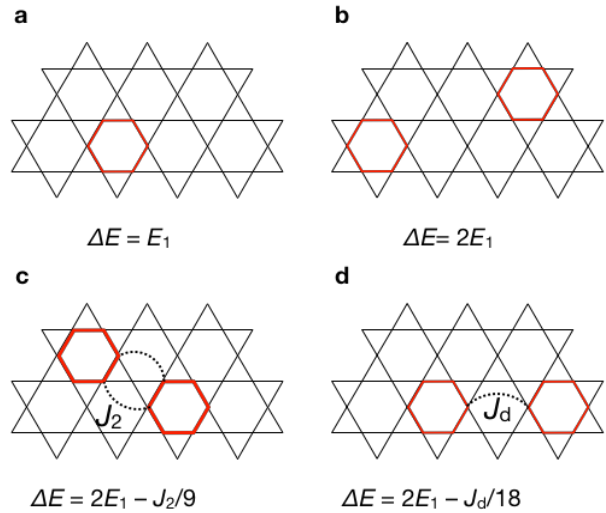

**S5 Comparison of energies for various magnon configurations.** We consider a $J$–$J_2$–$J_d$ model in a magnetic field larger than saturation, in which a fully polarized state is the ground state and regarded as a magnon vacuum with zero energy. **a**, the formation of a single hexagonal magnon with $S_z = 2$ costs $\Delta E = E_1$. **b**, $2E_1$ for two independent magnons without mutual interactions. **c,** configuration of two nearby magnons expected for $Q_{mag} = 9$, which are connected by $J_2$. $\Delta E$ is now $2E_1 - J_2/9$. **d,** configuration of two nearby magnons for $Q_{mag} = 12$ with $J_d$ coupling, resulting $\Delta E = 2E_1 - J_d/18$. This simple argument reveals that antiferromagnetic $J_2$ and $J_d$ stabilize magnon configurations with $Q_{mag} = 9$ and 12, respectively, while ferromagnetic interactions leads to destabilization. For CdK, all configurations with $Q_{mag} = 12$ are observed, while only on-site 3-magnon configuration with $S^z = 0$ is observed for $Q_{mag} = 9$. This fact strongly suggests ferromagnetic $J_2$ and antiferromagnetic $J_d$. Magnon crystals with larger unit cells in high fields have no energy gain or loss by $J_2$ or $J_d$ and are supposed to originate from higher order interactions. We have assigned 19/21 instead of 32/36 to $m_6$ and 10/12 instead of 30/36 to $m_5$, because adding another magnon to the same hexagon costs a large on-site repulsion energy on the order of $J$: it is preferable to increase the density of single magnons rather than generating double or triple magnons of less density.